\pdfoutput=1

\documentclass[10pt]{bmc_article}    

\usepackage{amssymb, amsmath}
\usepackage{graphicx}
\usepackage{paralist}

\usepackage{hyperref}
\usepackage{url}
\urldef{\mailsin}\path|sergey@logic.pdmi.ras.ru|
\urldef{\mailmax}\path|maxal@cse.sc.edu|

\usepackage{soul}
\usepackage{cite}
\usepackage{multirow}
\usepackage{tikz}
\usetikzlibrary{plotmarks,shapes,snakes}
\usepackage{pgfplots}
\usepackage{icomma}
\usepackage{url}
\usepackage{xspace}

\usepackage{algorithmicx}
\usepackage{algorithm}
\usepackage{algpseudocode}
\algnotext{EndFor}
\algnotext{EndIf}
\algnotext{EndLoop}
\algnotext{EndWhile}
\algnotext{EndProcedure}
\algnotext{EndFunction}
\algrenewcommand\algorithmicrequire{\textbf{Input:}}
\makeatletter
\newcommand{\StatexIndent}[1][3]{%
  \setlength\@tempdima{\algorithmicindent}%
  \Statex\hskip\dimexpr#1\@tempdima\relax}
\makeatother

\newcommand{\BayesHammer}{\textsc{BayesHammer}\xspace}
\newcommand{\Hammer}{\textsc{Hammer}\xspace}
\newcommand{\SPAdes}{\textsc{SPAdes}\xspace}
\newcommand{\Quake}{\textsc{Quake}\xspace}

\definecolor{lightblue}{RGB}{140,140,255}
\definecolor{lightred}{RGB}{255,100,100}
\definecolor{lightgreen}{RGB}{200,255,200}
\definecolor{darkgreen}{RGB}{100,155,100}

\tikzstyle{every node}=[font=\footnotesize]

\tikzstyle{gate}            = [circle,fill=white,draw=black,minimum size=8pt,minimum height=8pt,inner sep=2pt,font=\small]
\tikzstyle{block}           = [rectangle,fill=white,draw=black,minimum size=10pt,text width=200pt,inner sep=2pt,rounded corners,font=\small]
\tikzstyle{smallblock}      = [rectangle,fill=white,draw=black,minimum size=10pt,text width=80pt,inner sep=2pt,rounded corners,font=\small]
\tikzstyle{datablock}       = [rectangle,fill=lightblue,draw=black,minimum size=10pt,minimum width=120pt,inner sep=2pt,rounded corners,font=\small]
\tikzstyle{ifelse}          = [rectangle,fill=lightgreen,draw=black,minimum height=10pt,minimum width=100pt,inner sep=2pt,rounded corners,font=\small]
\tikzstyle{wire}            = [draw,thick,->]
\tikzstyle{ewire}           = [draw,thick]
\tikzstyle{edgelabel}       = [draw=white,fill=white,circle,inner sep=0pt]
\tikzstyle{enode}           = [rectangle,inner sep=2pt,rounded corners]

\newcommand\dbg[1]{\texttt{#1}}

\def\ra#1{\rotatebox{90}{\parbox{2cm}{#1}}}

\def\amax{\mathrm{arg\,max}}

\def\mysize{\scriptsize}

\def\ham{\mathrm{d}}
\def\HG{\textsc{HG}}
\def\cal{\mathcal}
\def\file{\mathrm{File}}
\def\true{\textsc{true}}
\def\false{\textsc{false}}

\newcommand{\greyc}[1]{{\color{black!30!white}{#1}}}
\newcommand{\redc}[1]{{\color{red!80!white}{#1}}}
\newcommand{\bluec}[1]{{\color{blue!80!white}{#1}}}
\newcommand{\greenc}[1]{{\color{green!50!black}{#1}}}

\usepackage{cite} 
\usepackage{url}  
\usepackage{ifthen}  
\usepackage{multicol}   
\usepackage[utf8]{inputenc} 
\usepackage{pxfonts}

\newboolean{publ}

\newenvironment{bmcformat}{\baselineskip20pt\sloppy\setboolean{publ}{false}}{\baselineskip20pt\sloppy}

\begin{document}
\begin{bmcformat}


\title{BayesHammer: Bayesian clustering for error correction in single-cell sequencing}
 

\author{Sergey I. Nikolenko\correspondingauthor$^1$%
         \email{Sergey I. Nikolenko\correspondingauthor - \href{mailto:sergey@logic.pdmi.ras.ru}{sergey@logic.pdmi.ras.ru}},
         Anton I. Korobeynikov$^{1,2}$,
         and Max A. Alekseyev$^{1,3}$%
      }


\address{%
    \iid(1)Algorithmic Biology Laboratory, Academic University, St. Petersburg, Russia\\
    \iid(2)St. Petersburg State University, Russia\\
    \iid(3)Department of Computer Science and Engineering, University of South Carolina, Columbia, SC, USA
}%

\maketitle


\begin{abstract}
Error correction of sequenced reads remains a difficult task, especially in single-cell sequencing projects with 
extremely non-uniform coverage. While existing error correction tools designed for standard (multi-cell) sequencing 
data usually come up short in single-cell sequencing projects, algorithms actually used for single-cell error 
correction have been so far very simplistic.

We introduce several novel algorithms based on Hamming graphs and Bayesian subclustering in our new error
correction tool \BayesHammer. While \BayesHammer was designed for single-cell sequencing, we demonstrate that it also
improves on existing error correction tools for multi-cell sequencing data while working much faster on real-life datasets.
We benchmark \BayesHammer on both $k$-mer counts and actual assembly results with the \SPAdes genome assembler.
\end{abstract}

\ifthenelse{\boolean{publ}}{\begin{multicols}{2}}{}


\section*{Background}


Single-cell sequencing \cite{Grindberg2011,ChitsazEtAl11} based on the Multiple Displacement Amplification (MDA)
technology~\cite{Grindberg2011,Ishoey2008} allows one to sequence genomes of important uncultivated
bacteria that until recently had been viewed as unamenable to genome sequencing. Existing metagenomic
approaches (aimed at genes rather than genomes) are clearly limited for studies of such bacteria despite the fact that they
represent the majority of species in such important studies as the Human Microbiome Project
\cite{Gill2006,Hamady2009} or discovery of new antibiotics-producing bacteria \cite{LV09}.

Single-cell sequencing datasets have extremely non-uniform coverage that may vary from ones to thousands along a single genome (Fig.~\ref{fig-covplot}).
For many existing error correction tools, most notably \Quake~\cite{KSS10}, uniform coverage is a prerequisite: in the case of non-uniform coverage 
they either do not work or produce poor results.

Error correction tools usually attempt to correct the set of $k$-character substrings of reads called \emph{$k$-mers} and then propagate corrections
to whole reads which are important to have for many assemblers. Error correction tools often employ a simple idea of discarding
rare $k$-mers, which obviously does not work in the case of non-uniform coverage.

Medvedev et al.~\cite{MSP11} recently presented a new approach to error correction for datasets with non-uniform
coverage. Their algorithm \Hammer makes use of the Hamming graph (hence the name) on $k$-mers
(vertices of the graph correspond to $k$-mers and edges connect pairs of $k$-mers with Hamming distance 
not exceeding a certain threshold). \Hammer employs a simple and fast clustering technique
based on selecting a \emph{central $k$-mer} in each connected component of the Hamming graph. Such central $k$-mers are assumed to be
error-free (i.e., they are assumed to actually appear in the genome), while the other $k$-mers from connected components are assumed to be 
erroneous instances of the corresponding central $k$-mers.
However, \Hammer may be overly simplistic: in connected components of large diameter or connected components with several $k$-mers of large multiplicities,
it is more reasonable to assume that there are two or more central $k$-mers (rather than one as in \Hammer). 
Biologically, such connected components may correspond to either \begin{inparaenum}[(1)] 
\item repeated regions with similar but not identical genomic sequences (\emph{repeats}) which would be bundled together by existing error correction 
tools (including \Hammer); or \item artificially united $k$-mers from distinct parts of the genome that just happen to be connected by a path in the 
Hamming graph (characteristic to \Hammer).\end{inparaenum} 

In this paper, we introduce the \BayesHammer error correction tool that does not rely on uniform coverage. \BayesHammer
uses the clustering algorithm of \Hammer as a first step and  then refines the constructed clusters by further 
subclustering them with a procedure that takes into account reads quality values (e.g., provided by Illumina sequencing 
machines) and introduces Bayesian (BIC) penalties for extra subclustering parameters. \BayesHammer subclustering aims to 
capture the complex structure of repeats (possibly of varying coverage) in the genome by separating even very similar 
$k$-mers that come from different instances of a repeat. \BayesHammer also uses a new approach for propagating 
corrections in $k$-mers to corrections in the reads. All algorithms in \BayesHammer are heavily parallelized 
whenever possible; as a result, \BayesHammer gains a significant speedup with more processing cores available.
These features make \BayesHammer a perfect error correction tool for single-cell sequencing. 

We remark that \Hammer produces only a set of central $k$-mers but does not correct reads, making it incompatible 
with most genome assemblers. \Quake does correct reads but has severe memory limitations for large $k$ and assumes 
uniform coverage. In contrast, \textsc{EULER-SR}~\cite{CP08} and \textsc{Camel}~\cite{ChitsazEtAl11} correct reads and 
do not make strong assumptions on coverage (both tools have been used for single-cell assembly projects 
\cite{ChitsazEtAl11}) which makes these tools suitable for comparison to \BayesHammer.
Our benchmarks show that \BayesHammer outperforms these tools in both single-cell and standard (multi-cell) modes. 
We further couple \BayesHammer with a recently developed genome assembler \SPAdes~\cite{Bankevich2012} and
demonstrate that assembly of \BayesHammer-corrected reads significantly
improves upon assembly with reads corrected by other tools for the same datasets, while the total running time also improves significantly.

\BayesHammer is freely available for download as part of the \SPAdes genome assembler
at \url{http://bioinf.spbau.ru/spades/}.

\begin{figure}[t]
\begin{center}
\begin{tikzpicture}[scale=1.0]
\begin{semilogyaxis}[xlabel={\scriptsize KBases},ylabel={},ymin=0,xmin=0,xmax=4639,height=3cm,width=\columnwidth, scaled x ticks=false,
x tick label style={font=\scriptsize}, y tick label style={font=\scriptsize} ]
\input{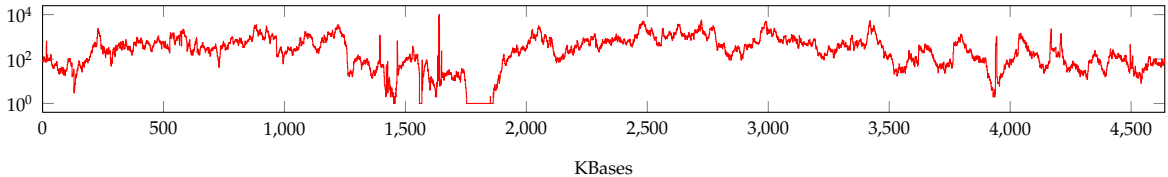}
\end{semilogyaxis}
\end{tikzpicture}
\end{center}
\caption{Logarithmic coverage plot for the single-cell \emph{E. coli} dataset (similar plot is also given in \cite{ChitsazEtAl11}).}\label{fig-covplot}
\end{figure}

\section*{Methods}

\subsection*{Notation and outline}

Let $\Sigma=\{A, C, G, T\}$ be the alphabet of nucleotides (\BayesHammer discards $k$-mers with uncertain bases denoted $N$).
A $k$-mer is an element of $\Sigma^k$, i.e., a string of $k$ nucleotides.
We denote the $i^{\texttt{th}}$ letter (nucleotide) of a $k$-mer $x$ by $x[i]$, indexing them from
zero: $0\leq i\leq k-1$.
A subsequence of $x$ corresponding to a set of indices $I$ is denoted by $x[I]$.
We use interval notation $[i, j]$ for intervals of integers $\{i, i+1, \ldots, j\}$ and further abbreviate
$x[i, j]=x\left[\{i, i+1, \ldots, j\}\right]$;
thus, $x=x[0, k-1]$. Input reads are represented as a set of strings $R\subset\Sigma^*$ along with their
{\em quality values} $(q_r[i])_{i=0}^{|r|-1}$ for each $r\in R$.
We assume that $q_r[i]$ estimates the probability that there has been an
error in position $i$ of read $r$. Notice that in practice, the \texttt{fastq} file format \cite{Cock2010} contains
characters that encode probabilities on a logarithmic scale (in particular, products of probabilities used below
correspond to sums of actual quality values).

Below we give an overview of \BayesHammer workflow (Fig.~\ref{fig-workflow}) and refer to subsequent sections for further details.
On Step~(1), $k$-mers in the reads are counted,
producing a triple $statistics(x)=(count_x, quality_x, \textbf{error}_x)$  for each $k$-mer $x$. Here,
$count_x$ is the number of times $x$ appears as a substring in the reads,
$quality_x$ is its total quality expressed as a probability of sequencing error in $x$, and $\textbf{error}_x$ is a $k$-dimensional vector
that contains products of error probabilities (sums of quality values) for individual nucleotides of $x$
across all its occurrences in the reads. 
On Step~(2), we find connected components of the Hamming graph constructed from this set of $k$-mers.
On Step~(3), the connected components become subject to Bayesian subclustering;  
as a result, for each $k$-mer we know the center of its subcluster. On Step~(4), we filter subcluster
centers according to their total quality and form a set of \emph{solid} $k$-mers which is then iteratively expanded on Step~(5) by mapping them back
to the reads. 
Step~(6) deals with reads correction by counting the majority vote of solid
$k$-mers in each read. 
In the iterative version, if there has been a substantial amount of changes in the
reads, we run the next iteration of error correction; otherwise, output the corrected reads.
Below we describe specific algorithms employed in the \BayesHammer pipeline.

\begin{figure}[t]
{
\begin{center}
\begin{tikzpicture}[line width=0.5pt, every node/.style = {anchor = base}, xscale=0.8, yscale=0.7]
  \begin{scope}
    \node[datablock] (init) at (10,11) {\mysize Set of reads};
    \node[block]  (counts) at (10,10) {\mysize (1) Compute $k$-mer statistics from reads};
    \draw[wire] (init) -- (counts);
    \node[block]  (hammer) at (10,9) {\mysize (2) Construct connected components of Hamming graph};
    \draw[wire] (counts) -- (hammer);
    \node[block]  (bayes) at (10,8) {\mysize (3) Bayesian subclustering of the connected components};
    \draw[wire] (hammer) -- (bayes);
    \node[block]  (filter) at (10,7) {\mysize (4) Select solid $k$-mers from subcluster centers};
    \draw[wire] (bayes) -- (filter);
    \node[datablock]  (solid) at (10,5.7) {\mysize A set of solid $k$-mers};
    \draw[wire] (filter) -- (solid);
    \node[block]  (iterative) at (10,5) {\mysize (5) Iteratively expand the set of solid $k$-mers};
    \draw[wire] (solid) -- (iterative);
    \node[ifelse]  (iterres) at (10,3.7) {\mysize Set of solid $k$-mers changed?};
    \draw[wire] (iterative) -- (iterres);
    \draw[wire] (iterres)  -- node[near start,above] {\mysize yes} ++(4.75cm,0) |-  (iterative);

    \node[block]  (reconstruct) at (10,2.8) {\mysize (6) Correct reads};
    \draw[wire] (iterres) edge node[left] {\mysize no} (reconstruct);

    \node[ifelse] (finalif) at (10,1.5) {\mysize Reads substantially changed?};
    \node[datablock] (results) at (10,0.5) {\mysize A set of corrected reads};
    \draw[wire] (reconstruct) -- (finalif);
    \draw[wire] (finalif) -- node[above, near start] {\mysize yes} ++(4.95cm,0) |- (counts);
    \draw[wire] (finalif) edge node[left] {\mysize no} (results);
  \end{scope}
\end{tikzpicture}
\end{center}
}

\caption{\BayesHammer workflow.}\label{fig-workflow}
\end{figure}

\subsection*{Algorithms}

\subsubsection*{Step~(1): computing $k$-mer statistics.}\label{sub-count}

To collect $k$-mer statistics, we use a straightforward hash map approach \cite{CLR09} that does not require 
storing instances of all $k$-mers in memory (as excessive amount of RAM might be needed otherwise).
For a certain positive integer $N$ (the number of auxiliary files), we use a hash function $h:\Sigma^k\to\mathbb Z_N$
that maps $k$-mers over the alphabet $\Sigma$ to integers from $0$ to $N-1$.

\begin{algorithm}
\caption{Count $k$-mers}\label{step1}
\begin{algorithmic}[0]
\For{each $k$-mer $x$ from the reads $R$:}
  \State compute $h(x)$ and write $x$ to $\file_{h(x)}$.
\EndFor
\For{$i\in[0, N-1]$:}
  \State sort $\file_i$ with respect to the lexicographic order;
  \State reading $\file_i$ sequentially, compute $statistics(s)$ for each $k$-mer $s$ from $\file_i$.
\EndFor
\end{algorithmic}
\end{algorithm}


\subsubsection*{Step~(2): Constructing connected components of Hamming graph.}\label{sub-hammer}

Step~(2) is the essence of the \Hammer approach \cite{MSP11}.
The \emph{Hamming distance} between $k$-mers $x,y\in\Sigma^k$ 
is the number of nucleotides in which they differ:
$$\ham(x, y) = \left|\{ i\in[0,k-1]\,:\, x[i] \neq y[i] \}\right|.$$
For a set of $k$-mers $X$, 
the \emph{Hamming graph}
$\HG_\tau(X)$ is an undirected graph with the set of vertices $X$ and edges corresponding 
to pairs of $k$-mers from $X$ with Hamming distance at most $\tau$, i.e., $x, y\in X$ are connected by an edge in $\HG_\tau(X)$ iff $\ham(x, y)\le\tau$ (Fig.~\ref{fig-restricted}).
To construct $\HG_\tau(X)$ efficiently, we notice that if two $k$-mers are at Hamming distance at most $\tau$, and we partition
the set of indices $[0, k-1]$ into $\tau+1$ parts, then at least one part corresponds to the same subsequence in both $k$-mers.
Below we assume with little loss of generality that $\tau+1$ divides $k$, i.e., $k = \sigma\cdot (\tau+1)$ for some integer $\sigma$.

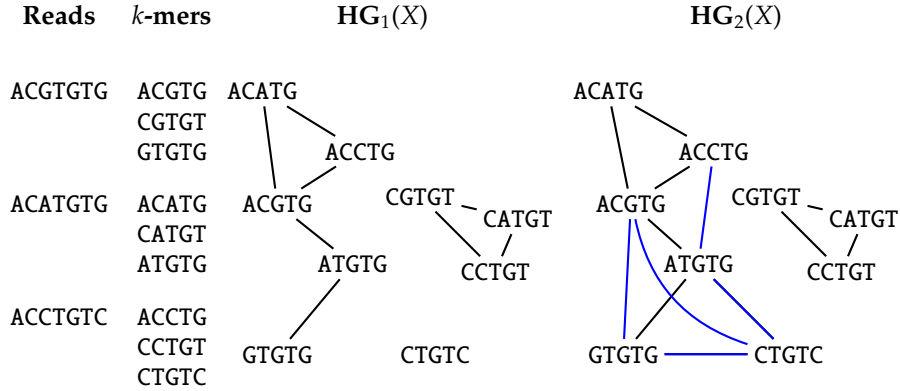
\begin{figure}[t]

{
\begin{center} 
\begin{tikzpicture}[scale=1, line width=1pt, every node/.style = {anchor = base}]
  \begin{scope}
	\node[enode] (reads) at (1.5, 6) { \textbf{Reads} };
	\node[enode] (read1) at (1.5, 5) { \dbg{ACGTGTG} };
	\node[enode] (read2) at (1.5, 3.5) { \dbg{ACATGTG} };
	\node[enode] (read3) at (1.5, 2) { \dbg{ACCTGTC} };

	\node[enode] (kmers) at (3, 6) { \textbf{$k$-mers} };
	\node[enode] (kmer11) at (3, 5) { \dbg{ACGTG} };
	\node[enode] (kmer12) at (3, 4.6) { \dbg{CGTGT} };
	\node[enode] (kmer13) at (3, 4.2) { \dbg{GTGTG} };

	\node[enode] (kmer21) at (3, 3.5) { \dbg{ACATG} };
	\node[enode] (kmer22) at (3, 3.1) { \dbg{CATGT} };
	\node[enode] (kmer23) at (3, 2.7) { \dbg{ATGTG} };

	\node[enode] (kmer31) at (3, 2) { \dbg{ACCTG} };
	\node[enode] (kmer32) at (3, 1.6) { \dbg{CCTGT} };
	\node[enode] (kmer33) at (3, 1.2) { \dbg{CTGTC} };

	\node[enode] (kmers) at (5.8, 6) { \textbf{$\HG_1(X)$} };
	\node[enode] (k11) at (4.4, 3.5) { \dbg{ACGTG} };
	\node[enode] (k12) at (6.3, 3.6) { \dbg{CGTGT} };
	\node[enode] (k13) at (4.4, 1.5) { \dbg{GTGTG} };

	\node[enode] (k21) at (4.2, 5) { \dbg{ACATG} };
	\node[enode] (k22) at (7.6, 3.3) { \dbg{CATGT} };
	\node[enode] (k23) at (5.4, 2.7) { \dbg{ATGTG} };

	\node[enode] (k31) at (5.5, 4.2) { \dbg{ACCTG} };
	\node[enode] (k32) at (7.3, 2.6) { \dbg{CCTGT} };
	\node[enode] (k33) at (6.5, 1.5) { \dbg{CTGTC} };

	\draw[ewire] (k11) -- (k21);
	\draw[ewire] (k11) -- (k31);
	\draw[ewire] (k21) -- (k31);
	\draw[ewire] (k11) -- (k23);

	\draw[ewire] (k13) -- (k23);
	\draw[ewire] (k12) -- (k22);
	\draw[ewire] (k32) -- (k22);
	\draw[ewire] (k12) -- (k32);

	\node[enode] (kmers) at (10.5, 6) { \textbf{$\HG_2(X)$} };
	\node[enode] (k11) at (9.1, 3.5) { \dbg{ACGTG} };
	\node[enode] (k12) at (10.9, 3.6) { \dbg{CGTGT} };
	\node[enode] (k13) at (9.0, 1.5) { \dbg{GTGTG} };

	\node[enode] (k21) at (8.8, 5) { \dbg{ACATG} };
	\node[enode] (k22) at (12.2, 3.3) { \dbg{CATGT} };
	\node[enode] (k23) at (10.0, 2.7) { \dbg{ATGTG} };

	\node[enode] (k31) at (10.2, 4.2) { \dbg{ACCTG} };
	\node[enode] (k32) at (11.9, 2.6) { \dbg{CCTGT} };
	\node[enode] (k33) at (11.2, 1.5) { \dbg{CTGTC} };

	\draw[ewire] (k11) -- (k21);
	\draw[ewire] (k11) -- (k31);
	\draw[ewire] (k21) -- (k31);
	\draw[ewire] (k11) -- (k23);
	\draw[ewire] (k13) -- (k23);
	\draw[ewire] (k33) -- (k23);
	\draw[ewire] (k13) -- (k33);
	\draw[ewire] (k12) -- (k22);
	\draw[ewire] (k32) -- (k22);
	\draw[ewire] (k12) -- (k32);

	\draw[ewire,thick,color=blue] (k31) -- (k23);
	\draw[ewire,thick,color=blue] (k11) -- (k13);
	\draw[ewire,bend right=30,thick,color=blue] (k11) edge (k33);
	\draw[ewire,thick,color=blue] (k33) -- (k23);
	\draw[ewire,thick,color=blue] (k13) -- (k33);

  \end{scope}
\end{tikzpicture}
\end{center}
}

\caption{Hamming graphs $\HG_1(X)$ and $\HG_2(X)$ for $X$ being the set of 4-mers 
of the reads $\dbg{ACGTGTG}$, $\dbg{ACATGTG}$, $\dbg{ACCTGTC}$. Blue edges denote Hamming distance $2$.}
\label{fig-restricted}
\end{figure}

For a subset of indices $I\subseteq [0, k-1]$, we define a partial lexicographic ordering $\prec_I$
as follows: $x\prec_I y$ iff $x[I]\prec y[I]$, where $\prec$ is the lexicographic ordering on $\Sigma^*$.
Similarly, we define a partial equality $=_I$ such that $x=_I y$ iff $x[I]=y[I]$.
We partition the set of indices $[0, k-1]$ into $\tau+1$ parts of size $\sigma$
and for each part $I$, sort a separate copy of $X$ with respect to $\prec_I$.
As noticed above, for every two $k$-mers $x, y\in X$ with $\ham(x, y)\le\tau$, there exists a part $I$ such that $x=_I y$. 
It therefore suffices to separately consider blocks of equivalent $k$-mers with respect to $=_I$ for each part $I$.
If a block is small (i.e., of size smaller than a certain threshold), we go over the pairs of $k$-mers in this block to find
those with Hamming distance at most $\tau$.
If a block is large, we recursively apply to it the same procedure with a different partition of the indices.
In practice, we use two different partitions of $[0, k-1]$: the first corresponds to contigious subsets of indices 
(recall that $\sigma=\tfrac{k}{\tau+1}$):
$$I^{\textrm{cnt}}_s=\{s\sigma, s\sigma+1, \ldots, s\sigma+\sigma-1\}, \quad s=0, \ldots, \tau,$$
while the second corresponds to strided subsets of indices:
$$I^{\textrm{str}}_s=\{s, s+\tau+1, s+2(\tau+1), \ldots, s+(\sigma-1)(\tau+1)\}, \quad s=0, \ldots, \tau.$$
\BayesHammer uses a two-step procedure, first splitting with respect to $\{I^{\textrm{cnt}}_s\}_{s=0}^{\tau}$ 
(Fig.~\ref{fig-subkmers}) and then, if an equivalence block is large, with respect to $\{I^{\textrm{str}}_s\}_{s=0}^{\tau}$.
On the block processing step, we use the disjoint set data structure \cite{CLR09} to maintain the set
of connected components. Step~(2) is summarized in Algorithm~\ref{step2}.

\begin{algorithm}[t]
\caption{Hamming graph processing}\label{step2}
\begin{algorithmic}[0]
\Procedure{HGProcess}{$X$, max\_quadratic}
  \State Init components with singletons $\cal X = \{ \{x\}: x\in X\}$.
  \ForAll{$Y\in\mathrm{FindBlocks}(X, \{I^{\textrm{cnt}}_s\}_{s=0}^{\tau})$}
    \If{$|Y| > \texttt{max\_quadratic}$}
      \ForAll{$Z\in\mathrm{FindBlocks}(Y, \{I^{\textrm{str}}_s\}_{s=0}^{\tau})$}
        \State $\mathrm{ProcessExhaustively}(Z, \cal X)$
      \EndFor
    \Else
      \State $\mathrm{ProcessExhaustively}(Y, \cal X)$.
    \EndIf
  \EndFor
\EndProcedure
\Function{FindBlocks}{$X$, $\{I_s\}_{s=0}^{\tau}$}
  \For{$s=0, \ldots, \tau$}
    \State sort a copy of $X$ with respect to $\prec_{I_s}$, getting $X_s$.
  \EndFor
  \For{$s=0, \ldots, \tau$}
    \State output the set of equiv. blocks $\{Y\}$ w.r.t. $=_{I_s}$.
  \EndFor
\EndFunction
\Procedure{ProcessExhaustively}{$Y$, $\cal X$}
  \For{each pair $x, y\in Y$}
    \If{$\ham(x, y)\le\tau$} join their sets in $\cal X$:
      \ForAll{$x\in Z_x\in\cal X$, $y\in Z_y\in\cal X$}
        \State $\cal X:=\cal X\cup\{ Z_x\cup Z_y \}\setminus \{Z_x, Z_y\}$.
      \EndFor
    \EndIf
  \EndFor
\EndProcedure
\end{algorithmic}
\end{algorithm}

\begin{figure}[t]
{
\begin{center} 
\begin{tikzpicture}[scale=1.0, line width=1pt, every node/.style = {anchor = base}]
  \begin{scope}
	\node[enode] (1kmers) at (3, 6) { \textbf{$X$} };
	\node[enode] (1kmer1) at (3, 5)   { \verb$1 $\bluec{\tt ACG}\greenc{\tt TGT}\redc{\tt GTA} };
	\node[enode] (1kmer2) at (3, 4.6) { \verb$2 $\bluec{\tt CGT}\greenc{\tt GTG}\redc{\tt TAA} };
	\node[enode] (1kmer3) at (3, 4.2) { \verb$3 $\bluec{\tt GTG}\greenc{\tt TGT}\redc{\tt AAC} };
	\node[enode] (1kmer4) at (3, 3.8)   { \verb$4 $\bluec{\tt ACC}\greenc{\tt TGT}\redc{\tt GTA} };
	\node[enode] (1kmer5) at (3, 3.4)   { \verb$5 $\bluec{\tt CCT}\greenc{\tt GTG}\redc{\tt TAC} };
	\node[enode] (1kmer6) at (3, 3)     { \verb$6 $\bluec{\tt CTG}\greenc{\tt TGT}\redc{\tt ACT} };

	\node[enode] (2kmers) at (5, 6) { \textbf{ $X[I^{\textrm{cnt}}_0]$} };
	\node[enode] (2kmer1) at (5, 5)   { \verb$1 $\bluec{\tt ACG} };
	\node[enode] (2kmer2) at (5, 4.6) { \verb$2 $\bluec{\tt CGT} };
	\node[enode] (2kmer3) at (5, 4.2) { \verb$3 $\bluec{\tt GTG} };
	\node[enode] (2kmer4) at (5, 3.8)   { \verb$4 $\bluec{\tt ACC} };
	\node[enode] (2kmer5) at (5, 3.4) { \verb$5 $\bluec{\tt CCT} };
	\node[enode] (2kmer6) at (5, 3) { \verb$6 $\bluec{\tt CTG} };

	\node[enode] (3kmers) at (6.5, 6) { \textbf{ $X[I^{\textrm{cnt}}_1]$} };
	\node[enode] (3kmer1) at (6.5, 5)   { \verb$1 $\greenc{\tt TGT} };
	\node[enode] (3kmer2) at (6.5, 4.6) { \verb$2 $\greenc{\tt GTG} };
	\node[enode] (3kmer3) at (6.5, 4.2) { \verb$3 $\greenc{\tt TGT} };
	\node[enode] (3kmer4) at (6.5, 3.8) { \verb$4 $\greenc{\tt TGT} };
	\node[enode] (3kmer5) at (6.5, 3.4) { \verb$5 $\greenc{\tt GTG} };
	\node[enode] (3kmer6) at (6.5, 3)   { \verb$6 $\greenc{\tt TGT} };

	\node[enode] (3akmers) at (8, 6) { \textbf{ $X[I^{\textrm{cnt}}_2]$} };
	\node[enode] (3akmer1) at (8, 5)   { \verb$1 $\redc{\tt GTA} };
	\node[enode] (3akmer2) at (8, 4.6) { \verb$2 $\redc{\tt TAA} };
	\node[enode] (3akmer3) at (8, 4.2) { \verb$3 $\redc{\tt AAC} };
	\node[enode] (3akmer4) at (8, 3.8) { \verb$4 $\redc{\tt GTA} };
	\node[enode] (3akmer5) at (8, 3.4) { \verb$5 $\redc{\tt TAC} };
	\node[enode] (3akmer6) at (8, 3)   { \verb$6 $\redc{\tt ACT} };

	\node[enode] (4kmers) at (9.5, 6) { \textbf{ $X[I^{\textrm{cnt}}_0]$} };
	\node[enode] (4kmerss) at (9.5, 5.5) { \textbf{ sorted } };
	\node[enode] (4kmer1) at (9.5, 4.6) { \verb$1 $\bluec{\tt ACG} };
	\node[enode] (4kmer2) at (9.5, 3.8) { \verb$2 $\bluec{\tt CGT} };
	\node[enode] (4kmer3) at (9.5, 3)   { \verb$3 $\bluec{\tt GTG} };
	\node[enode] (4kmer4) at (9.5, 5)   { \verb$4 $\bluec{\tt ACC} };
	\node[enode] (4kmer5) at (9.5, 4.2) { \verb$5 $\bluec{\tt CCT} };
	\node[enode] (4kmer6) at (9.5, 3.4) { \verb$6 $\bluec{\tt CTG} };

	\node[enode] (5kmers) at (11, 6) { \textbf{ $X[I^{\textrm{cnt}}_1]$} };
	\node[enode] (5kmerss) at (11, 5.5) { \textbf{ sorted } };
	\node[enode] (5kmer1) at (11, 4.2)   { \verb$1 $\greenc{\tt TGT} };
	\node[enode] (5kmer2) at (11, 5) { \verb$2 $\greenc{\tt GTG} };
	\node[enode] (5kmer3) at (11, 3.8) { \verb$3 $\greenc{\tt TGT} };
	\node[enode] (5kmer4) at (11, 3.4)   { \verb$4 $\greenc{\tt TGT} };
	\node[enode] (5kmer5) at (11, 4.6) { \verb$5 $\greenc{\tt GTG} };
	\node[enode] (5kmer6) at (11, 3) { \verb$6 $\greenc{\tt TGT} };

	\node[enode] (3bkmers) at (12.5, 6) { \textbf{ $X[I^{\textrm{cnt}}_2]$} };
	\node[enode] (5bkmerss) at (12.5, 5.5) { \textbf{ sorted } };
	\node[enode] (3bkmer1) at (12.5, 4.2)   { \verb$1 $\redc{\tt GTA} };
	\node[enode] (3bkmer2) at (12.5, 3.4) { \verb$2 $\redc{\tt TAA} };
	\node[enode] (3bkmer3) at (12.5, 5) { \verb$3 $\redc{\tt AAC} };
	\node[enode] (3bkmer4) at (12.5, 3.8) { \verb$4 $\redc{\tt GTA} };
	\node[enode] (3bkmer5) at (12.5, 3) { \verb$5 $\redc{\tt TAC} };
	\node[enode] (3bkmer6) at (12.5, 4.6)   { \verb$6 $\redc{\tt ACT} };
	
	\draw[red,thick,dotted] (10.3,5.3) -- (10.3, 4.55) -- (11.7, 4.55) -- (11.7, 5.3) -- (10.3,5.3);
	\draw[red,thick,dotted] (10.3,4.5) -- (10.3, 2.9) -- (11.7, 2.9) -- (11.7, 4.5) -- (10.3,4.5);
	\draw[red,thick,dotted] (11.8,4.5) -- (11.8, 3.7) -- (13.2, 3.7) -- (13.2, 4.5) -- (11.8,4.5);
  \end{scope}
\end{tikzpicture}
\end{center}
}

\caption{Partial lexicographic orderings of a set $X$ of $9$-mers with respect to the index sets
$I^{\textrm{cnt}}_0 = \{ 0,1,2\}$, $I^{\textrm{cnt}}_1=\{3,4,5\}$, and $I^{\textrm{cnt}}_2=\{6,7,8\}$.
Red dotted lines indicate equivalence blocks.}
\label{fig-subkmers}
\end{figure}

\subsubsection*{Step~(3): Bayesian subclustering.}\label{sub-bayes}

In \Hammer's generative model~\cite{MSP11}, it is assumed that errors 
in each position of a $k$-mer are independent and occur with the same probability $\epsilon$, which is a fixed global 
parameter (\Hammer used $\epsilon=0.01$). Thus, the likelihood that 
a $k$-mer $x$
was generated from 
a $k$-mer $y$
under \Hammer's model equals
$$
L_{\Hammer}(x\mid y) = (1-\epsilon)^{k-\ham(x, y)}\epsilon^{\ham(x, y)}.
$$
Under this model, the maximum likelihood center of a cluster is simply its consensus string~\cite{MSP11}.

In \BayesHammer, we further elaborate upon \Hammer's model. Instead of a fixed $\epsilon$, we use reads quality values 
that approximate probabilities $q_x[i]$ of a nucleotide at position $i$ in the $k$-mer $x$ being erroneous.
We combine quality values from identical $k$-mers in the reads: for a multiset
of $k$-mers $X$ that agree on the $j^{\textrm{th}}$ nucleotide, it is erroneous with probability $\prod_{x\in X}q_x[j]$. 

The likelihood that a $k$-mer $x$ has been generated from another $k$-mer $c$ (under the independent errors assumption)
is given by 
$$L(x\mid c)=\prod_{j:\ x[j]\neq c[j]}q_{x}[j]\prod_{j:\ x[j]=c[j]}\left(1-q_{x}[j]\right),$$
and the likelihood of a specific subclustering $C=C_1\cup\ldots\cup C_m$ is
$$L_m(C_1,\ldots,C_m)=\prod_{i=1}^m \prod_{x\in C_i} L(x\mid c_i)$$
where $c_i$ is the center (consensus string) of the subcluster $C_i$. 

In the subclustering procedure (see Algorithm~\ref{step3}), we sequentially subcluster each connected component of
the Hamming graph into more and more clusters with the classical $k$-means clustering algorithm (denoted $m$-means
since $k$ has different meaning). For the objective function, we use the likelihood as above penalized for overfitting with
the Bayesian  information criterion (BIC) \cite{Schwarz78}. In this case, there are
$|C|$ observations in the dataset, and the total number of parameters is $3km+m-1$:
\begin{itemize}
\item $m-1$ for probabilities of subclusters,
\item $km$ for cluster centers, and
\item $2km$ for error probabilities in each letter: there are $3$ possible errors for each letter, and the probabilities
should sum up to one. Here error probabilities are conditioned on the fact that an error has occurred (alternatively,
we could consider the entire distribution, including the correct letter, and get $3km$ parameters for probabilities
but then there would be no need to specify cluster centers, so the total number is the same).
\end{itemize}

Therefore, the resulting objective function is $$\ell_m := 2\cdot \log L_m(C_1,\ldots,C_m) - (3km + m-1)\cdot \log |C|$$
for subclustering into $m$ clusters; we stop as soon as $\ell_m$ ceases to increase.

\begin{algorithm}[t]
\caption{Bayesian subclustering}\label{step3}
\begin{algorithmic}[0]
\ForAll{connected components $C$ of the Hamming graph}
  \State $m := 1$
  \State $\ell_1 := 2\log L_1(C)$ (likelihood of the cluster generated by the consensus)
  \Repeat
    \State $m := m+1$
    \State do $m$-means clustering of $C=C_1\cup\ldots\cup C_m$ w.r.t. the Hamming 
  distance; the initial approximation
  \StatexIndent[2]  to the centers is given by $k$-mers that have the least error probability
  	\State $\ell_m := 2\cdot \log L_m(C_1,\ldots,C_m) - (3km + m-1)\cdot \log |C|$
  \Until{$\ell_m\le\ell_{m-1}$}
  \State output the best found clustering $C=C_1\cup\ldots\cup C_{m-1}$
\EndFor
\end{algorithmic}
\end{algorithm}

\subsubsection*{Steps~(4) and~(5): selecting solid $k$-mers and expanding the set of solid $k$-mers.}\label{sub-filter}

We define the quality of a $k$-mer $x$ as the probability that it is error-free: $p_{x} = \prod_{j=0}^{k-1}\left(1-q_x[j]\right).$
The $k$-mer qualities are computed on Step~(1) along with computing $k$-mer statistics. Next,
we (generously) define the quality of a cluster $C$ as the probability that at least one $k$-mer in $C$ is correct:
$$p_C = 1 - \prod_{x\in C}\left(1-p_x\right).$$
In contrast to \Hammer, we do not distinguish whether the cluster
is a singleton (i.e., $|C|=1$); there may be plenty of superfluous clusters with several $k$-mers obtained by chance
(actually, it is more likely to obtain a cluster of several $k$-mers by chance than a singleton of the same total multiplicity).

Initially we mark as \emph{solid} the centers of the clusters whose total quality exceeds a predefined threshold (a global parameter for \BayesHammer,
set to be rather strict).
Then we expand the set of solid $k$-mers iteratively:
if a read is completely covered by solid $k$-mers we conclude that it actually comes from the genome and mark all other $k$-mers in this
read as solid, too (Algorithm~\ref{step5}).

\begin{algorithm}[t]
\caption{Solid $k$-mers expansion}\label{step5}
\begin{algorithmic}[0]
\Procedure{IterativeExpansion}{$R$, $X$}
\While{$\mathrm{ExpansionStep}(R, X)$}
\EndWhile
\EndProcedure
\Function{ExpansionStep}{$R$, $X$}
\ForAll{reads $r\in R$}
  \If{$r$ is completely covered by solid $k$-mers}
    \State mark all $k$-mers in $r$ as solid
  \EndIf
\EndFor
\State Return $\true$ if $X$ has increased and $\false$ otherwise.
\EndFunction
\end{algorithmic}
\end{algorithm}

\subsubsection*{Step~(6): reads correction.}\label{sub-reconstruct}

After Steps~(1)-(5), we have constructed the set of solid $k$-mers that are presumably error-free.
To construct corrected reads from the set of solid $k$-mers, for each base of every read, we compute
the consensus of all solid $k$-mers and solid centers of clusters of all non-solid $k$-mers covering this base (Fig.~\ref{fig-reconstruct}).
This step is formally described as Algorithm~\ref{step6}.

\begin{algorithm}[t]
\caption{Reads correction}\label{step6}
\begin{algorithmic}[0]
\Require reads\,$R$,\,solid\,$k$-mers\,$X$,\,clusters\,$\mathcal C$.
\ForAll{reads $r\in R$}
  \State init consensus array $v:[0, |r|-1]\times \{A, C, G, T\}\to\mathbb N$
  with zeros: $v(j, x[i]) := 0$ for all $i=0,\dots,|r|-1$ and $j=0,\dots,k-1$
  \For{$i=0, \ldots, |r|-k$}
  	\If{$r[i, i+k-1]\in X$ (it is solid)}
      \For{$j\in[i, i+k-1]$}
        \State $v(j, r[i]) := v(j, r[i]) + 1$
      \EndFor
    \EndIf
    \If{$r[i, i+k-1]\in C$ for some $C\in\mathcal C$}
      \State let $x$ be the center of $C$
      \If{$x\in X$ ($r$ belongs to a cluster with solid center)}
      	\For{$j\in[i, i+k-1]$}
	      \State $v(j, x[i]) := v(j, x[i]) + 1$
      	\EndFor
      \EndIf
    \EndIf
  \EndFor
  \For{$i\in[0, |r|-1]$}
    \State $r[i] := \amax_{a\in \Sigma} v(i, a)$.
  \EndFor
\EndFor
\end{algorithmic}
\end{algorithm}

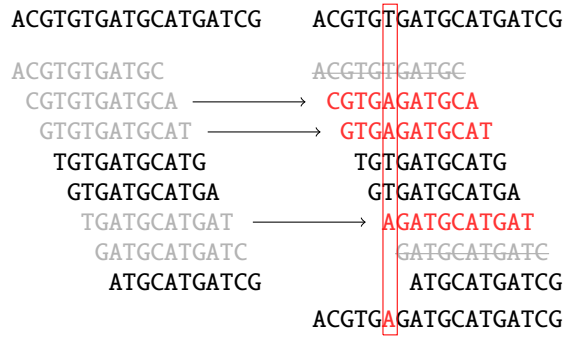
\begin{figure}[t]
{
\begin{center}
\hspace{-1cm}
\begin{tikzpicture}[line width=1pt, every node/.style = {anchor = base}]
  \begin{scope}

	\node[enode] (read)  at (3, 5.3) { {\tt \phantom{AAAAAAA}ACGTGTGATGCATGATCG } };
	\node[enode] (kmer1) at (3, 4.6) {\greyc{\tt ACGTGTGATGC       } };
	\node[enode] (kmer2) at (3, 4.2) {\greyc{\tt \phantom{AA}CGTGTGATGCA      } };
	\node[enode] (kmer3) at (3, 3.8) {\greyc{\tt \phantom{AAAA}GTGTGATGCAT     } };
	\node[enode] (kmer4) at (3, 3.4) {		 {\tt \phantom{AAAAAA}TGTGATGCATG    } };
	\node[enode] (kmer5) at (3, 3.0) { 		{\tt \phantom{AAAAAAAA}GTGATGCATGA   } };
	\node[enode] (kmer6) at (3, 2.6) {\greyc{\tt \phantom{AAAAAAAAAA}TGATGCATGAT  } };
	\node[enode] (kmer7) at (3, 2.2) {\greyc{\tt \phantom{AAAAAAAAAAAA}GATGCATGATC } };
	\node[enode] (kmer8) at (3, 1.8) { 		{\tt \phantom{AAAAAAAAAAAAA}ATGCATGATCG} };

	\node[enode] (read) at (7, 5.3) { {\tt \phantom{AAAAAAA}ACGTGTGATGCATGATCG } };
	\node[enode] (kmer21) at (7, 4.6) {\greyc{\tt \st{ACGTGTGATGC}       } };
	\node[enode] (kmer22) at (7, 4.2) {\redc{\tt \phantom{AA}CGTGAGATGCA      } };
	\node[enode] (kmer23) at (7, 3.8) {\redc{\tt \phantom{AAAA}GTGAGATGCAT     } };
	\node[enode] (kmer24) at (7, 3.4) {		 {\tt \phantom{AAAAAA}TGTGATGCATG    } };
	\node[enode] (kmer25) at (7, 3.0) { 		{\tt \phantom{AAAAAAAA}GTGATGCATGA   } };
	\node[enode] (kmer26) at (7, 2.6) {\redc{\tt \phantom{AAAAAAAAAA}AGATGCATGAT  } };
	\node[enode] (kmer27) at (7, 2.2) {\greyc{\tt \phantom{AAAAAAAAAAAA}\st{GATGCATGATC} } };
	\node[enode] (kmer28) at (7, 1.8) { 		{\tt \phantom{AAAAAAAAAAAAA}ATGCATGATCG} };

	\node[enode] (read) at (7, 1.3) { {\tt \phantom{AAAAAAA}ACGTG\redc{A}GATGCATGATCG } };
\draw[->,thin] (4.3,4.3) -- (5.8, 4.3);
\draw[->,thin] (4.5,3.9) -- (6.0, 3.9);
\draw[->,thin] (5.1,2.7) -- (6.6, 2.7);

\def\x{6.83}
\def\y{1.2}
\def\yy{5.6}
\def\xx{7.02}

\draw[red,thin] (\x,\y) -- (\x,\yy) -- (\xx,\yy) -- (\xx,\y) -- (\x,\y);


  \end{scope}
\end{tikzpicture}
\end{center}
}

\caption{Reads correction. Grey $k$-mers indicate non-solid $k$-mers. Red $k$-mers are the centers of the corresponding
clusters (two grey $k$-mers striked through on the right are non-solid singletons). As a result, one nucleotide is changed.}
\label{fig-reconstruct}
\end{figure}

\section*{Results and discussion}

\subsection*{Datasets} 
In our experiments, we used three datasets from \cite{ChitsazEtAl11}: a single-cell \emph{E. coli}, a single-cell \emph{S. aureus},
and a standard (multicell) \emph{E. coli} dataset. Paired-end libraries were generated by an Illumina Genome
Analyzer IIx from MDA-amplified single-cell DNA and from multicell genomic DNA prepared from cultured \emph{E. coli}, respectively
These datasets consist of $100$bp paired-end reads with insert size $220$; both \emph{E. coli} datasets have average coverage $\approx 600\times$, although the
coverage is highly non-uniform in the single-cell case.

In all experiments, \BayesHammer used $k=21$ (we observed no improvements for higher values of $k$).

\subsection*{$k$-mer counts}

\begin{table*}[t] \footnotesize
\begin{center}
\begin{tabular}{|l|p{25pt}|r|r|r|p{50pt}|p{50pt}|p{40pt}|}\hline
Correction tool & Running time & \multicolumn{5}{c|}{$k$-mers} & Reads \\\hline
 & & Total & Genomic & Non-genomic & \% of all genomic $k$-mers found in reads & \% genomic among all $k$-mers in reads & \% reads aligned to genome \\\hline
 & & \multicolumn{6}{|c|}{ {\bf Multi-cell \emph{E. coli}}, total $4{,}543{,}849$ genomic $k$-mers} \\\hline
Uncorrected       &       & $187{,}580{,}875$ & $4{,}543{,}684$ & $183{,}037{,}191$ & $99.99$ & $2.4$ & $99.05$ \\
Quake             &       &   $4{,}565{,}237$ & $4{,}543{,}461$ &        $21{,}776$ & $99.99$ & $99.5$ & $99.97$ \\
HammerNoExpansion & $30$m &  $58{,}305{,}738$ & $4{,}543{,}674$ &  $53{,}762{,}064$ & $99.99$ & $8.4$ & $95.59$ \\
HammerExpanded    & $36$m &  $28{,}290{,}788$ & $4{,}543{,}673$ &  $23{,}747{,}115$ & $99.99$ & $19.1$ & $99.49$ \\
BayesHammer       & $37$m &  $27{,}100{,}305$ & $4{,}543{,}674$ &  $22{,}556{,}631$ & $99.99$ & $20.1$ & $99.62$ \\\hline
& & \multicolumn{6}{|c|}{ {\bf Single-cell \emph{E. coli}}, total $4{,}543{,}849$ genomic $k$-mers} \\\hline
Uncorrected       &           & $165{,}355{,}467$ & $4{,}450{,}489$ & $160{,}904{,}978$ & $97.9$ & $2.7$ & $79.05$ \\
Camel             & $2$h$29$m & $147{,}297{,}070$ & $4{,}450{,}311$ & $142{,}846{,}759$ & $97.9$ & $3.0$ & $81.25$ \\
Euler-SR          & $2$h$15$m & $138{,}677{,}818$ & $4{,}450{,}431$ & $134{,}227{,}387$ & $97.9$ & $3.2$ & $81.95$ \\
Coral             & $2$h$47$m & $156{,}907{,}496$ & $4{,}449{,}560$ & $152{,}457{,}936$ & $97.9$ & $2.8$ & $80.28$ \\
HammerNoExpansion & $37$m     &  $53{,}001{,}778$ & $4{,}443{,}538$ &  $48{,}558{,}240$ & $97.8$ & $8.3$ & $81.36$ \\
HammerExpanded    & $43$m     &  $36{,}471{,}268$ & $4{,}443{,}545$ &  $32{,}027{,}723$ & $97.8$ & $12.1$ & $86.91$ \\
BayesHammer       & $57$m     &  $35{,}862{,}329$ & $4{,}443{,}736$ &  $31{,}418{,}593$ & $97.8$ & $12.4$ & $87.12$ \\\hline
& & \multicolumn{6}{|c|}{ {\bf Single-cell \emph{S. aureus}}, total $2{,}821{,}095$ genomic $k$-mers} \\\hline
Uncorrected       &           & $88{,}331{,}311$ & $2{,}820{,}394$ & $85{,}510{,}917$ & $99.98$ & $3.2$ & $75.07$ \\
Camel             & $5$h$13$m & $69{,}365{,}311$ & $2{,}820{,}350$ & $66{,}544{,}961$ & $99.97$ & $4.1$ & $75.27$ \\
Euler-SR          & $2$h$33$m & $58{,}886{,}372$ & $2{,}820{,}349$ & $56{,}066{,}023$ & $99.97$ & $4.8$ & $75.24$ \\
Coral             & $7$h$12$m & $83{,}249{,}146$ & $2{,}820{,}011$ & $80{,}429{,}135$ & $99.96$ & $3.4$ & $75.22$ \\
HammerNoExpansion & $58$m     & $37{,}465{,}296$ & $2{,}820{,}341$ & $34{,}644{,}955$ & $99.97$ & $7.5$ & $71.63$ \\
HammerExpanded    & $1$h$03$m & $23{,}197{,}521$ & $2{,}820{,}316$ & $20{,}377{,}205$ & $99.97$ & $12.1$ & $76.54$ \\
BayesHammer       & $1$h$09$m & $22{,}457{,}509$ & $2{,}820{,}311$ & $19{,}637{,}198$ & $99.97$ & $12.6$ & $76.60$ \\
\hline
\end{tabular}
\vskip5pt
\caption{$k$-mer statistics.}
\label{tbl:kmers}
\end{center}
\end{table*}

Table~\ref{tbl:kmers} shows error correction statistics produced by different tools on all three datasets. For a comparison with \Hammer,
we have emulated \Hammer with read correction by turning off Bayesian subclustering (\emph{HammerExpanded} in the table) and both
Bayesian subclustering and read expansion, another new idea of \BayesHammer (\emph{HammerNoExpansion} in the table).
Note that despite its more complex processing, \BayesHammer is significantly faster than other error correction tools
(except, of course, for \Hammer which is a strict subset of \BayesHammer processing in our experiments and is run on \BayesHammer code).
\BayesHammer also produces, in the single-cell case, a much smaller set of $k$-mers in the resulting reads which
leads to smaller de Bruijn graphs and thus reduces the total assembly running time.
Since \BayesHammer trims only bad quality bases and does not, like \Quake, trim bases that it has not been able to correct
(it has been proven detrimental for single-cell assembly in our experiments), it does produce a much larger set of $k$-mers than
Quake on a multi-cell dataset.

For a comparison of \BayesHammer with other tools in terms of error rate reduction across an average read, see
the logarithmic error rate graphs on Fig.~\ref{fig:bherrors}. Note that we are able to count errors only for the
reads that actually aligned to the genome, so the graphs are biased in this way. Note how the first $21$ bases are corrected better
than others in \BayesHammer and both versions of \Hammer since we have run it with $k=21$; still, other values of $k$ did not show
a significant improvement in either $k$-mer statistics or, more importantly, assembly results.

\begin{figure*}[t]
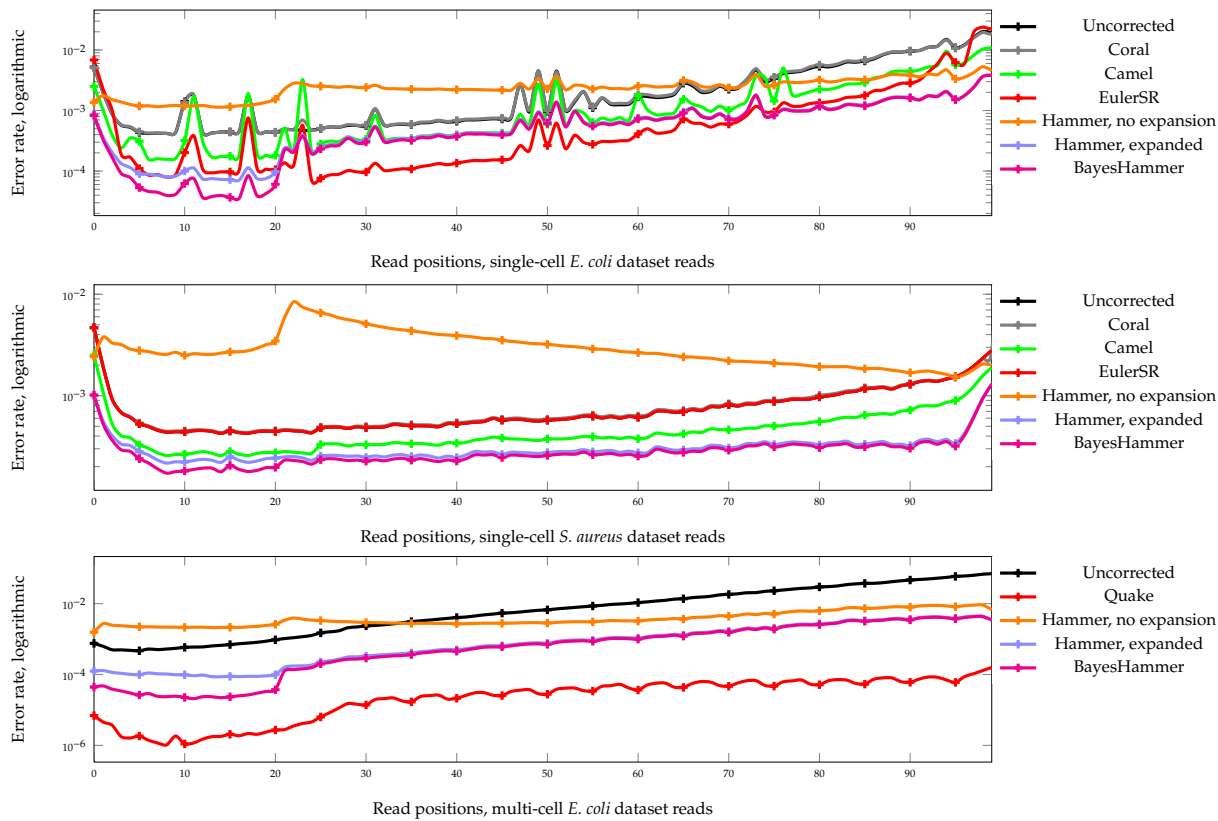

\begin{center}
\begin{tikzpicture}[scale=0.8]
\begin{semilogyaxis}[xlabel={Read positions, single-cell \emph{E. coli} dataset reads},ylabel={Error rate, logarithmic},height=5cm,width=\textwidth, scaled x ticks=true, scaled y ticks=true,
legend style={legend columns=1,at={(1,1)},anchor=north west,font=\tiny,draw=none}, xmin=0, xmax=99,
x tick label style={font=\tiny}, y tick label style={font=\tiny} ]
\input{covgraphs/ecoli.uncorr.single.log.cov.errors.tex}
\input{covgraphs/ecoli.coral.single.log.cov.errors.tex}
\input{covgraphs/ecoli.camel.log.cov.errors.tex}
\input{covgraphs/ecoli.euler.log.cov.errors.tex}
\input{covgraphs/ecoli.hammernoext.single.log.cov.errors.tex}
\input{covgraphs/ecoli.hammer.single.log.cov.errors.tex}
\input{covgraphs/ecoli.bh.single.log.cov.errors.tex}
\legend{ {Uncorrected}, {Coral}, {Camel}, {EulerSR}, {Hammer, no expansion}, {Hammer, expanded}, {BayesHammer} }
\end{semilogyaxis}
\end{tikzpicture}
\begin{tikzpicture}[scale=0.8]
\begin{semilogyaxis}[xlabel={Read positions, single-cell \emph{S. aureus} dataset reads},ylabel={Error rate, logarithmic},height=5cm,width=\textwidth, scaled x ticks=true, scaled y ticks=true,
legend style={legend columns=1,at={(1,1)},anchor=north west,font=\tiny,draw=none}, xmin=0, xmax=99,
x tick label style={font=\tiny}, y tick label style={font=\tiny} ]
\input{covgraphs/sau.uncorr.single.log.cov.errors.tex}
\input{covgraphs/sau.coral.log.cov.errors.tex}
\input{covgraphs/sau.camel.single.log.cov.errors.tex}
\input{covgraphs/sau.euler.log.cov.errors.tex}
\input{covgraphs/sau.hammernoext.single.log.cov.errors.tex}
\input{covgraphs/sau.hammer.single.log.cov.errors.tex}
\input{covgraphs/sau.bh.single.log.cov.errors.tex}
\legend{ {Uncorrected}, {Coral}, {Camel}, {EulerSR}, {Hammer, no expansion}, {Hammer, expanded}, {BayesHammer} }
\end{semilogyaxis}
\end{tikzpicture}
\begin{tikzpicture}[scale=0.8]
\begin{semilogyaxis}[xlabel={Read positions, multi-cell \emph{E. coli} dataset reads},ylabel={Error rate, logarithmic},height=5cm,width=\textwidth, scaled x ticks=true, scaled y ticks=true,
legend style={legend columns=1,at={(1,1)},anchor=north west,font=\tiny,draw=none}, xmin=0, xmax=99,
x tick label style={font=\tiny}, y tick label style={font=\tiny} ]
\input{covgraphs/mce.uncorr.log.cov.errors.tex}
\input{covgraphs/mce.quake.single.log.cov.errors.tex}
\input{covgraphs/mce.hammernoext.single.log.cov.errors.tex}
\input{covgraphs/mce.hammer.single.log.cov.errors.tex}
\input{covgraphs/mce.bh.single.log.cov.errors.tex}
\legend{ {Uncorrected}, {Quake}, {Hammer, no expansion}, {Hammer, expanded}, {BayesHammer} }
\end{semilogyaxis}
\end{tikzpicture}
\end{center}
\caption{Error reduction by read position on logarithmic scale for the single-cell \emph{E. coli},
single-cell \emph{S. aureus}, and multi-cell \emph{E. coli} datasets.}\label{fig:bherrors}
\end{figure*}

\subsection*{Assembly results}

\begin{table}[ht] \scriptsize
\begin{center}
\caption{Assembly results, single-cell \emph{E. coli} and \emph{S. aureus} datasets (contigs of length $\geq$ 200 are used).}\label{tbl:scres}
\begin{tabular}{|l*{10}{|p{26pt}}|}
\hline
Statistics & \ra{BayesHammer} & \ra{BayesHammer\newline (scaffold)} & \ra{Coral} & \ra{Coral (scaffold)} & \ra{EulerSR} & \ra{EulerSR (scaffold)} &
\ra{Hammer,\newline expanded} & \ra{Hammer,\newline no expansion} & \ra{Hammer,\newline no expansion\newline (scaffold)} &
\ra{Hammer\newline (scaffold)} \\ \hline
& \multicolumn{10}{c|}{ {\bf Single-cell \emph{E. coli}}, reference length $4639675$, reference GC content $50.79\%$ } \\\hline
\# contigs ($\geq$ 1000 bp) & 191 & 158 & 276 & 224 & 231 & 150 & 195 & 282 & 242 & 173 \\ 
\# contigs & 521 & 462 & 675 & 592 & 578 & 375 & 529 & 655 & 592 & 477 \\ 
Largest contig & 269177 & 284968 & 179022 & 179022 & 267676 & 267676 & 268464 & 210850 & 210850 & 268464 \\ 
Total length & 4952297 & 4989404 & 5064570 & 5122860 & 4817757 & 4902434 & 4977294 & 5097148 & 5340871 & 5005022 \\ 
N50 & 110539 & 113056 & 45672 & 67849 & 74139 & 95704 & 97639 & 65415 & 84893 & 109826 \\ 
NG50 & 112065 & 118432 & 55073 & 87317 & 77762 & 108976 & 101871 & 68595 & 96600 & 112161 \\ 
NA50 & 110539 & 113056 & 45672 & 67765 & 74139 & 95704 & 97639 & 65415 & 84841 & 109826 \\ 
NGA50 & 112064 & 118432 & 55073 & 87317 & 77762 & 108976 & 101871 & 68594 & 96361 & 112161 \\ 
\# misassemblies & 4 & 6 & 9 & 12 & 6 & 8 & 4 & 4 & 7 & 7 \\ 
\# misassembled contigs & 4 & 6 & 9 & 10 & 6 & 8 & 4 & 4 & 7 & 7 \\ 
Misass. contigs length & 42496 & 94172 & 62114 & 150232 & 47372 & 149639 & 43304 & 26872 & 147140 & 130706 \\ 
Genome covered (\%) & 96.320 & 96.315 & 96.623 & 96.646 & 95.337 & 95.231 & 96.287 & 96.247 & 96.228 & 96.281 \\ 
GC (\%) & 49.70 & 49.69 & 49.61 & 49.56 & 49.90 & 49.74 & 49.68 & 49.64 & 49.60 & 49.68 \\ 
\# mismatches / 100kbp & 11.22 & 11.70 & 8.36 & 9.10 & 5.55 & 5.82 & 12.77 & 54.11 & 52.48 & 13.08 \\ 
\# indels / 100kbp & 1.07 & 8.26 & 9.17 & 12.76 & 0.52 & 47.80 & 0.91 & 1.17 & 7.96 & 8.69 \\ 
\# genes & 4065 + 124 part & 4079 + 110 part & 3998 + 180 part & 4040 + 143 part & 3992 + 140 part & 4020 + 107 part & 4068 + 123 part & 4034 + 152 part & 4048 + 136 part & 4078 + 111 part \\ \hline
& \multicolumn{10}{c|}{ {\bf Single-cell \emph{S. aureus}}, reference length $2872769$, reference GC content $32.75\%$ } \\\hline
\# contigs ($\geq$ 1000 bp) & 95 & 85 & 132 & 113 & 82 & 70 & 114 & 272 & 258 & 101 \\ 
Total length ($\geq$ 1000 bp) & 3019597 & 3309342 & 3055585 & 3066662 & 2972925 & 2993100 & 3033912 & 3389846 & 3405223 & 3509555 \\ 
\# contigs & 260 & 241 & 455 & 423 & 166 & 134 & 312 & 721 & 711 & 292 \\ 
Largest contig & 282558 & 328686 & 208166 & 208166 & 254085 & 535477 & 282558 & 148002 & 166053 & 328679 \\ 
Total length & 3081173 & 3368034 & 3160497 & 3166169 & 3008746 & 3020256 & 3111423 & 3575679 & 3594468 & 3584266 \\ 
N50 & 87684 & 145466 & 62429 & 90701 & 101836 & 145466 & 74715 & 30788 & 34943 & 131272 \\ 
NG50 & 112566 & 194902 & 87636 & 99341 & 108151 & 159555 & 88292 & 39768 & 45889 & 180022 \\ 
NA50 & 87684 & 145466 & 62429 & 89365 & 100509 & 145466 & 68711 & 30788 & 34552 & 112801 \\ 
NGA50 & 88246 & 148064 & 74452 & 90101 & 101836 & 145466 & 88289 & 35998 & 42642 & 148023 \\ 
\# misassemblies & 15 & 17 & 11 & 14 & 4 & 5 & 11 & 14 & 18 & 14 \\ 
\# misassembled contigs & 12 & 14 & 9 & 10 & 4 & 5 & 9 & 14 & 16 & 12 \\ 
Misass. contigs length & 340603 & 779785 & 478009 & 523596 & 377133 & 918380 & 402997 & 272677 & 324361 & 940356 \\ 
Genome covered (\%) & 99.522 & 99.483 & 99.449 & 99.447 & 99.213 & 99.254 & 99.204 & 98.820 & 98.888 & 99.221 \\ 
GC (\%) & 32.67 & 32.63 & 32.64 & 32.63 & 32.66 & 32.67 & 32.67 & 32.39 & 32.38 & 32.57 \\ 
\# mismatches per 100 kbp & 3.18 & 8.01 & 12.44 & 12.65 & 9.72 & 10.28 & 17.38 & 54.92 & 55.50 & 15.36 \\ 
\# indels per 100 kbp & 2.17 & 2.30 & 15.50 & 15.67 & 3.80 & 4.08 & 3.57 & 2.64 & 2.72 & 3.04 \\ 
\# genes & 2540 + 36 part & 2547 + 30 part & 2532 + 45 part & 2540 + 37 part & 2547 + 30 part & 2550 + 27 part & 2535 + 41 part & 2477 + 91 part & 2485 + 85 part & 2539 + 38 part \\ \hline
\end{tabular}
\end{center}
\end{table}

\begin{table}[ht] \scriptsize
\begin{center}
\caption{Assembly results, multi-cell \emph{E. coli} dataset (contigs of length $\geq$ 200 are used).}\label{tbl:mcres}
\begin{tabular}{|l*{10}{|p{40pt}}|}
\hline
Statistics & \ra{BayesHammer} & \ra{BayesHammer\newline (scaffold)} & 
\ra{Hammer,\newline expanded} & \ra{Hammer,\newline no expansion} & \ra{Hammer,\newline no expansion\newline (scaffold)} &
\ra{Hammer\newline (scaffold)} & \ra{Quake} \\ \hline
& \multicolumn{7}{c|}{ {\bf Multi-cell \emph{E. coli}}, $600\times$ coverage, reference length $4639675$, reference GC content $50.79\%$ } \\\hline
\# contigs ($\geq$ 500 bp) & 103 & 102 & 119 & 238 & 213 & 115 & 165 \\ 
\# contigs ($\geq$ 1000 bp) & 91 & 90 & 99 & 192 & 171 & 96 & 156 \\ 
Total length ($\geq$ 500 bp) & 4641845 & 4641790 & 4626515 & 4730338 & 4817457 & 4627067 & 4543682 \\ 
Total length ($\geq$ 1000 bp) & 4633361 & 4633306 & 4611745 & 4696966 & 4787210 & 4612838 & 4537565 \\ 
\# contigs & 122 & 121 & 146 & 325 & 303 & 141 & 204 \\ 
Largest contig & 285113 & 285113 & 218217 & 210240 & 210240 & 218217 & 165487 \\ 
Total length & 4647325 & 4647270 & 4635156 & 4756088 & 4844208 & 4635349 & 4555015 \\ 
N50 & 132645 & 132645 & 113608 & 59167 & 73113 & 113608 & 58777 \\ 
NG50 & 132645 & 132645 & 113608 & 59669 & 80085 & 113608 & 57174 \\ 
NA50 & 132645 & 132645 & 113608 & 59167 & 73113 & 113608 & 58777 \\ 
NGA50 & 132645 & 132645 & 113608 & 59669 & 80085 & 113608 & 57174 \\ 
\# misassemblies & 3 & 3 & 4 & 4 & 7 & 5 & 0 \\ 
\# misassembled contigs & 3 & 3 & 4 & 4 & 7 & 5 & 0 \\ 
Misassembled contigs length & 44466 & 44466 & 57908 & 15259 & 30901 & 60418 & 0 \\ 
Genome covered (\%) & 99.440 & 99.440 & 99.383 & 98.891 & 98.925 & 99.385 & 98.747 \\ 
GC (\%) & 50.78 & 50.77 & 50.77 & 50.73 & 50.71 & 50.77 & 50.75 \\ 
N's (\%) & 0.00000 & 0.00000 & 0.00000 & 0.00000 & 0.00000 & 0.00000 & 0.00000 \\ 
\# mismatches per 100 kbp & 8.55 & 8.55 & 13.76 & 44.46 & 44.33 & 13.76 & 1.21 \\ 
\# indels per 100 kbp & 0.99 & 0.99 & 1.14 & 0.76 & 0.97 & 1.14 & 0.20 \\ 
\# genes & 4254+45 part & 4254+45 part & 4245+56 part & 4196+72 part & 4204+68 part & 4245+56 part & 4174+62 part \\ \hline
\end{tabular}
\end{center}
\end{table}

Tables~\ref{tbl:scres} and~\ref{tbl:mcres} shows assembly results by the recently developed \SPAdes
assembler~\cite{Bankevich2012}; \SPAdes was designed specifically for single-cell assembly, but has by now demonstrated
state-of-the-art results on multi-cell datasets as well.

In the tables, N50 is such length that contigs of that length or longer comprise $\ge\tfrac12$ of the assembly; 
NG50 is a metric similar to N50 but only taking into account contigs comprising (and aligning to) the reference genome; 
NA50 is a metric similar to N50 after breaking up misassembled contigs by their misassemblies. NGx and NAx metrics have a
more direct relevance to assembly quality than regular Nx metrics; our result tables have been produced by the recently
developed tool QUAST \cite{Gurevich2012}.

All assemblies have been done with
\SPAdes. The results show that after \BayesHammer correction, assembly results improve significantly, especially in the
single-cell \emph{E. coli} case; it is especially interesting to note that even in the multi-cell case, where
\BayesHammer loses to \Quake by $k$-mer statistics, assembly results actually improve over assemblies produced from
\Quake-corrected reads (including genome coverage and the number of genes).


\section*{Conclusions}
Single-cell sequencing presents novel challenges to error correction tools. In contrast to multi-cell datasets, for single-cell datasets, there is no
pretty distribution of $k$-mer multiplicities; one therefore has to work with $k$-mers on a one-by-one basis, considering each cluster of $k$-mers separately. 
In this work, we further developed the ideas of \Hammer from a Bayesian clustering perspective and presented a new tool
\BayesHammer that makes them practical and yields significant improvements over existing error correction tools.

There is further work to be done to make our underlying models closer to real life; for instance, one could learn
a non-uniform distribution of single nucleotide errors and plug it in our likelihood formulas. Another natural
improvement would be to try and rid the results of contamination by either human or some other DNA material;
we observed significant human DNA contamination in our single-cell dataset, so weeding it out might yield a
significant improvement. Finally, a new general approach that we are going to try in our further work deals with
the technique of \emph{minimizers} introduced by Roberts et al. \cite{RHHMY04}.
It may provide significant reduction in memory requirements and a possible approach to dealing with paired information.



\section*{Acknowledgements}
\ifthenelse{\boolean{publ}}{\small}{} We thank Pavel Pevzner for many fruitful
discussions on all stages of the project. We are also grateful to Andrei
Prjibelski and Alexei Gurevich for help with the experiments and to the
anonymous referees whose comments have benefited the paper greatly.  This work
was supported the Government of the Russian Federation, grant 11.G34.31.0018.
Work of the first author was also supported by the Russian Fund for Basic
Research grant 12-01-00450-a and the Russian Presidential Grant MK-6628.2012.1.
Work of the second author was additionally supported by the Russian Fund for
Basic Research grant 12-01-00747-a.
  

\newpage
{\ifthenelse{\boolean{publ}}{\footnotesize}{\small}
 \bibliographystyle{bmc_article}  
  \bibliography{assembly} }     


\ifthenelse{\boolean{publ}}{\end{multicols}}{}

\end{bmcformat}
\end{document}